\magnification=\magstep1
\hfuzz=3truept
\baselineskip=14truept
\font\tif=cmr10 scaled \magstep3

\rightline{PUPT-1607}
\vfil
\centerline{\tif D-brane recoil}
\vfil
\centerline{{\rm Vipul 
Periwal}\footnote{${}^\dagger$}{vipul@puhep1.princeton.edu}
and \O yvind Tafjord\footnote{${}^*$}{otafjord@phoenix.princeton.edu}}
\bigskip
\centerline{Department of Physics}\centerline{Princeton University
}\centerline{Princeton, New Jersey 08544-0708}
\vfil
\par\noindent It is shown that there are logarithmic operators in
D-brane backgrounds that lead to infrared divergences in open string loop 
amplitudes.  These divergences can be cancelled by
changing the closed string background by operators that correspond to
the D-brane moving with constant velocity after some instant in time,
since it is precisely such operators that give rise to the appropriate
ultraviolet divergences in the closed string channel.
\medskip
\vfil
%\centerline{PACS: 11.17.+y}
%\vskip 1 truein %
\footline={\hfil}
\vfil\eject
\footline={\hss\tenrm\folio\hss}

\def\part{\partial}

\def\ws{1}
\def\as{2}
\def\jp{3}
\def\cf{4}
\def\fs{5}
\def\f{6}
\def\k{7}
\def\g{8}
Duality symmetries of various types have been of great interest in
recent times.  Such symmetries may lead to insights
into non-perturbative aspects of string theory, perhaps leading to
contact with observed physics.  This hope rests to some extent on
the celebrated work of Seiberg and Witten[\ws] on supersymmetric Yang-Mills
theories.  These symmetries predict the existence of solitonic objects in
certain vacua which are dual to `fundamental' objects in the dual vacua.
These solitonic objects play vital roles in a complete understanding 
of low-energy physics, as in the work of Strominger[\as]---it 
is therefore important to understand their dynamical properties.

Polchinski[\jp] showed 
that Dirichlet boundary conditions in open string theories,
previously studied for possible relevance to hadronic applications of
string physics, can be interpreted as an operational definition of 
solitons which carry Ramond-Ramond charge in closed string theories.  
These boundary conditions amount to an amazingly simple exact description
of solitons (called D-branes)
which are quite complicated from the spacetime perspective, and
are therefore a good laboratory for the study of dynamical properties of
solitons in string theory[\cf].  

In the present work, we will use a variant of the Fischler-Susskind[\fs] 
mechanism to treat the problem of D-brane recoil.  The problem of 
soliton recoil in field theory is already somewhat non-trivial, since 
one has to isolate the contributions of the Goldstone modes that
arise from broken symmetries, such as translation invariance.  In string
theory, it is not immediately clear how one could isolate such modes
in a consistent manner, especially from an exact description of the
string theory soliton[\cf].  The problem of soliton recoil in string theory
in a general case has been studied by Fischler, Paban and Rozali[\f], 
and in following work by Kogan and Mavromatos[\k].  
Our method will be much closer to Ref.~\fs, and our conclusions will
differ from Ref.'s \f,\k.  We will, however, confirm a `postulate' of
Kogan and Mavromatos concerning logarithmic operators[\g] in soliton 
backgrounds in string theory.

All the interesting elements of the problem are already evident in the
case of a 0-brane, which is just a particle from a spacetime perspective.
We will even restrict ourselves to the case of the bosonic string, since
no further conceptual light is shed on this problem by considering the
supersymmetric strings.  Of course, these solitonic objects are only
stable in supersymmetric theories, so an extension of our calculation
to such cases is definitely of interest.

The bosonic static 0-brane is described by imposing Dirichlet
boundary conditions on the spatial coordinates of the string, while
keeping Neumann boundary conditions on the time-like coordinate.  Thus
$$ X^i({\rm boundary}) = 0, \qquad \part_n X^0({\rm boundary}) =0,$$
where $\part_n$ is the derivative normal to the boundary of the string 
worldsheet, 
describes a 0-brane located at $x^i=0.$  Such a configuration obviously
breaks translation invariance in the spatial directions, and there
are vertex operators $V^i \equiv \oint \part_n X^i$ that translate the 0-brane
which correspond to Goldstone modes.  These vertex operators, however,
correspond to translations of the entire world-line of the 0-brane, and
cannot be used directly to describe recoil.  There are infrared divergences
in annulus amplitudes (which correspond to an open-string loop correction 
to the disk amplitude), and we shall see that they come from operators that
are rather closely related to these vertex operators.

Recall that the Fischler-Susskind mechanism[\fs] cancels infrared closed string
divergences due to massless dilatons at the one-loop level with a 
cosmological constant on the sphere.  We are interested in an infrared
divergence in an open string channel at the annulus level, 
which we aim to cancel with an
ultraviolet divergence in a closed string channel at the disk level.
\def\tr{{\rm tr}}
\def\ln{{\rm ln}}

To this end, we first calculate the annulus amplitude 
describing the scattering of one closed string tachyon off a 0-brane.
This calculation can be done in a variety of ways.  The simplest is
perhaps the operator formulation, in which case we need to calculate
$\tr V(k_1)\Delta V(k_2)\Delta,$ with $\Delta^{-1}\equiv L_0-1,$ and
$V(k_i)$ are the closed string vertex operators, integrated across the
propagating open string.  This calculation can also be formulated in
the closed string channel as
$\langle B|\Delta V(k_1)\Delta V(k_2) \Delta|B\rangle,$ where $|B\rangle$
is a state in the closed string Fock space that imposes the appropriate
boundary conditions on the end of the closed string worldsheet.
In either case, these calculations are uninteresting in themselves, as
far as the oscillator parts of the contractions are concerned---they give
the standard $\eta$ function form of the determinant of the Laplacian
on the annulus.  What is 
interesting, is the zero-mode trace, which we discuss in detail:
Writing 
$$\Delta \equiv \int_0^1 dx x^{L_0-2}, \qquad L_0 = 2p^2 +N,$$
the zero-mode trace for a 0-brane is
$$\int {dq^0\over2\pi} \langle q^0|\exp(-ik_1^0x^0) x_1^{-2(p^0)^2}
\exp(-ik_2^0x^0) x_2^{-2(p^0)^2}|q^0\rangle.$$
For our purposes, the important point is the dependence on $x_i,$
since we are interested in cancelling divergences that arise from 
$x_1\rightarrow 0,$ with $x_2$ held fixed (and vice versa).  This is
the limit when the annulus amplitude degenerates into a disk amplitude
with two open string insertions on the boundary.  The zero-mode trace
gives in this limit
$$ \delta(k^0_1+k^0_2) \sqrt{1\over \ln(x_1)}f(x_2,k_2^0).$$
This is the most important point of our calculation---the 0-brane
background implies that the zero-mode trace has changed from the 
standard open string, which has $26$ zero-mode integrals, giving
a factor of $ \ln(x_1)^{-13}.$
Since $x_1<1,$ there is an assumed analytic continuation, or an
$i\epsilon$ prescription in performing the Gaussian integral.
We shall find that precisely the same analytic continuation is needed
for the ultraviolet divergence we will find below, so the manner in
which one chooses to define the Gaussian integral is irrelevant,
provided it is chosen consistently.

Now, in the complete amplitude, we have, in the limit $x_1\rightarrow0,$
(neglecting divergences due to the pathologies of the bosonic open string
which have no dependence on the momenta of the closed string vertex
operators, and hence no bearing on the problem of recoil)
$$g_{{\rm st}}\int_{x_1\approx 0} 
{dx_1\over x_1\sqrt{8\pi\ln(x_1)}} A_{\rm disk}(k_1,k_2),$$
where 
$$ A_{\rm disk} = \langle V(k_1) V(k_2) V^i V^i\rangle.$$
We have found a rather peculiar feature, the divergence in the integral
over $x_1$ is proportional to $\sqrt{|\ln(\epsilon)|},$ where 
$-\ln(\epsilon)$ is the large-time infrared cutoff.  This must come from
open string states which have a two point function of the
form $\langle \phi(x) \phi(0)\rangle = \ln(x)/x^2.$   The 
operators $V^i$ are garden variety conformal fields, not capable of such
behaviour.  The appearance of logarithmic operators in
string backgrounds corresponding to solitons was postulated by
Kogan and Mavromatos[\k].  We have therefore confirmed their conjecture.  
We will explicitly find these operators,
which have a simple geometric interpretation, in the following.

Getting back to the matter at hand, we wish to cancel this divergence  
with a change in the closed string background.  We are looking for 
a closed string vertex operator that will lead to an ultraviolet 
divergence in a closed string channel, which would therefore be 
equivalent to an infrared divergence in an open string channel.
One might na\"\i vely think that one would need {\it two} open string
insertions, since the coefficient of the divergence is $ A_{\rm disk},$
but here we need to recall that the operators $V^i$ are very special
operators, since they produce infinitesimal motions of the entire 0-brane
worldline.  The effect of inserting any number of these operators can
be directly shown to be the same as multiplying the amplitude with 
factors of the total momentum carried by the external vertex operators.
We therefore consider a closed string vertex operator of the form
$$ V_{\rm recoil} \equiv \alpha^i \int d^2z \part_\alpha(f(X^0)\part^\alpha
X^i),$$
with $f$ to be determined.  By construction, such an 
operator gives a contribution only from boundary terms on the
worldsheet (which is the disk or the upper half plane), but it must be
regulated near the boundary because of the expectation value of $f(X^0).$
In fact, since the tangential derivative of $X^i$ at the boundary vanishes,
and energy is conserved at the order to which we are working, an
insertion of $V_{\rm recoil}$ is the same as an insertion of $\alpha^iV^i,$
but multiplied by this divergent expectation value.

Consider $f(X^0) \equiv  \int{(dq/2\pi)}\exp(iqX^0) g(q^2).$
(We assume that $f(X^0)$ has been normal-ordered on the sphere.)
We have, when $f(X^0)(z)$ is close to the boundary of the upper half plane,
$$\langle f(X^0)(i\epsilon/2)\rangle
 = \int{dq\over 2\pi}g(q^2)(\epsilon)^{-q^2}.$$
When $g(q^2) = 1/q^2,$ this gives the dependence on $\epsilon$ that
we need to cancel the divergence coming from the annulus.  It is, however, 
considerably more illuminating to write 
$$f(X^0) = X^0\Theta(X^0),$$
where $\Theta$ is the step function.  Thus, we have derived exactly what
we would na\"\i vely have predicted: The deformation of the D-brane 
background is precisely such that the D-brane starts moving at some time with 
constant velocity.  What is this velocity?  By comparing the annulus
divergence with the disk divergence due to $V_{\rm recoil},$ we
find
$$\alpha^i = 8\pi \sqrt2 g_{{\rm st}} (k_1+k_2)^i.$$
Recall that $\alpha^i X^0$ is the position of the soliton, 
and the mass of the 0-brane is expected $\propto 1/g_{{\rm st}},$ 
so this is exactly what we expect in soliton recoil.
One could treat this as a leading order determination of the mass of
the 0-brane.

In summary, we have arrived at a pleasing picture of 
D-brane recoil:  We have found logarithmic operators in the 
annulus amplitude, as conjectured in Ref.~\k.  We have cancelled
divergences in the disk amplitude due to insertions of $V_{\rm recoil}$
against divergences in the annulus amplitude due to the logarithmic operators,
in an ultraviolet$\leftrightarrow$infrared reversal of the 
Fischler-Susskind mechanism.  
The form of $V_{\rm recoil}$ we found has a simple and manifestly correct
physical meaning.  The next step is to extend these computations to
the physical case of the supersymmetric strings, and to compactifications 
which give rise to other finite mass D-branes.  At higher orders, one
could also derive energy conservation, though that likely involves including
other operators, such as $\oint \part_\tau X^0,$ along with
$V_{\rm recoil}.$  Presumably, higher order corrections will smooth out
the abrupt change in the soliton trajectory we have found at the leading order.
Carrying out the same
calculation for $p$-branes in flat space gives no divergences for $p>1,$ and
only a $\ln\ln \epsilon$ divergence for 1-branes, presumably related to
the properties of massless scalar fields on the worldsheet of the
1-brane.

\bigskip

Conversations with Larus Thorlacius are gratefully acknowledged.
This work was supported in part by NSF Grant No. PHY90-21984.
\hfuzz=2pt
\bigskip
\centerline{References}
\bigskip
\item{\ws} N. Seiberg and E. Witten, Nucl. Phys. B426:19-52(1994)
\item{\as} A. Strominger, Nucl. Phys. B451:96-108(1995)
\item{\jp} J. Polchinski, Phys. Rev. Lett. 75:4724-4727 (1995); for earlier
references, see J. Polchinski, S. Chaudhuri, C.V. Johnson,
{\it `Notes on D-branes'}, ITP Santa Barbara report NSF-ITP-96-003,
hep-th/9602052
\item{\cf} C. Callan and A. Felce, {\it `Soliton mass in string theory'},
unpublished report
\item{\fs} W. Fischler and L. Susskind, Phys. Lett. B171:383 (1986).
\item{\f} W. Fischler, S. Paban and M. Rozali, Phys. Lett. B352:298-303 (1995) 
\item{\k} I. Kogan and N. Mavromatos, {\it `Workdsheet logarithmic operators
and target space symmetries in string theory'}, Oxford report OUTP-95-50-P,
hep-th/9512210 
\item{\g} V. Gurarie, Nucl. Phys. B410:535 (1993)
\end